\newif\ifblind
\blindfalse

\documentclass{llncs}

\usepackage{graphicx}
\usepackage{booktabs}
\usepackage{url}

\usepackage{breakurl}
\usepackage[breaklinks,hidelinks]{hyperref}
\graphicspath{{figs/}}
\DeclareGraphicsExtensions{.png,.pdf}

\begin{document}

\date{}

\title{\Large \bf A Note on the Misinterpretation of the US Census Re-identification Attack}
\titlerunning{US Census Misinterpretation}

\ifblind
\author{
First Last, First Last, First Last \\
Affiliations \\
emails
}
\else
\author{Paul Francis}
\institute{Max Planck Institute for Software Systems (MPI-SWS)}
\fi

\maketitle

% Use the following at camera-ready time to suppress page numbers.
% Comment it out when you first submit the paper for review.
%\thispagestyle{empty}

%\subsection*{Abstract}

\begin{abstract}
 
In 2018, the US Census Bureau designed a new data reconstruction and re-identification attack and tested it against their 2010 data release. The specific attack executed by the Bureau allows an attacker to infer the race and ethnicity of respondents with average 75\% precision for 85\% of the respondents, assuming that the attacker knows the correct age, sex, and address of the respondents. They interpreted the attack as exceeding the Bureau's privacy standards, and so introduced stronger privacy protections for the 2020 Census in the form of the TopDown Algorithm (TDA).

This paper demonstrates that race and ethnicity can be inferred \emph{from the TDA-protected census data} with substantially \emph{better} precision and recall, using \emph{less} prior knowledge: only the respondents' address. Race and ethnicity can be inferred with average 75\% precision for 98\% of the respondents, and can be inferred with 100\% precision for 11\% of the respondents. The inference is done by simply assuming that the race/ethnicity of the respondent is that of the majority race/ethnicity for the respondent's census block.

We argue that the conclusion to draw from this simple demonstration is NOT that the Bureau's data releases lack adequate privacy protections. Indeed it is the Bureau's stated purpose of the data releases to allow this kind of inference. The problem, rather, is that the Bureau's criteria for measuring privacy is flawed and overly pessimistic. There is no compelling evidence that TDA was necessary in the first place.

\end{abstract}  

\section{Introduction}
\label{sec:intro}

The US Census Bureau releases privacy-protected statistics from the decennial census. In past decades, this data was protected using aggregation and swapping: occasionally exchanging an individual response from one geographic area, or block, with that in another block.

\begin{figure}
\begin{center}
\includegraphics[width=0.7\linewidth]{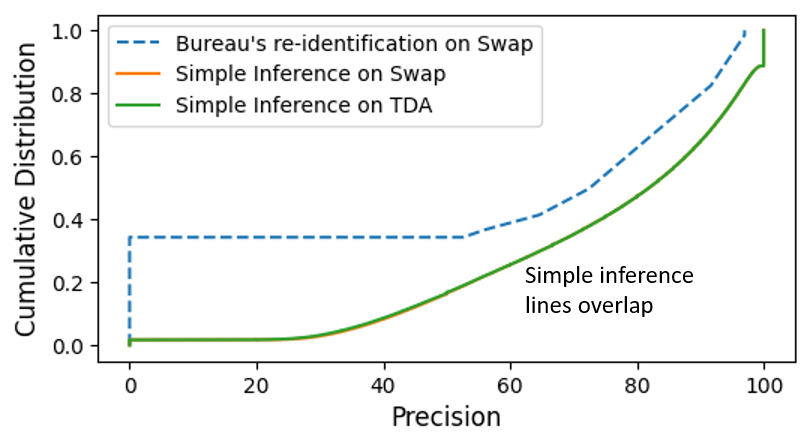}
\caption{CDF of precision for our simple inference ``non-attack'' run on the US Census Bureau's new disclosure protection mechanism (TDA) as well as the Bureau's prior mechanism (Swap). For comparison, the precision for the Bureau's re-identification attack on the prior swap mechanism is also shown. Points to the lower-right mean a more effective ``attack''. Note that effectiveness of the non-attack on swap-protected data is virtually identical to that of the TDA-protected data (lines overlap).}
\label{fig:precision_cdf}
\end{center}
\end{figure}

In 2019, the US Census Bureau reported on a new re-identification attack, developed by the Bureau, against these traditional \emph{swap-protected} data releases~\cite{abowd-2019-staring}. The attack was demonstrated on the 2010 release. The Bureau considered the attack serious enough that they developed a new privacy protection method. Called the Top-Down Algorithm (TDA), the new method uses aggregation and noise addition: perturbing counts with random noise from a normal distribution~\cite{census-disclosure-avoidance}. The 2020 census release is \emph{TDA-protected}.

The Bureau also prepared a TDA-protected release of the 2010 census so that stakeholders could evaluate data quality\footnote{
\href{https://www.census.gov/programs-surveys/decennial-census/decade/2020/planning-management/process/disclosure-avoidance/2020-das-development.html}{https://www.census.gov/programs-surveys/decennial-census/decade/2020/planning-management/process/disclosure-avoidance/2020-das-development.html}
}.

The specific re-identification attack demonstrated by the Bureau has two parts. First they reconstruct the original data from the swap-protected data. The reconstructed attributes are block, age, sex, race, and ethnicity (Hispanic or not). Next they link externally-derived data (address, age, and sex) with the reconstructed data to infer race and ethnicity of the re-identified respondents.  The re-identification attack achieved 75\% precision (75\% of race/ethnicity inferences were correct assuming correct prior knowledge).

This paper demonstrates that race/ethnicity inferences with \emph{better precision and recall} can be made against the 2010 \emph{TDA-protected release} using \emph{less prior knowledge} (only address instead of address, age, and sex). Our demonstration yields better than 95\% precision for 23\% of respondents, and virtually 100\% precision for 11\% of respondents (see Table~\ref{tab:overview}).

\begin{table*}
\centering
\begin{tabular}{r|ll}
  \toprule
    & Bureau's re-identification & Our simple inference \\
    & attack & ``non-attack'' \\
  \midrule
  Released data  & 2010 swap-protected release & 2010 TDA-protected release \\
  Prior knowledge & Address, age, sex & Address \\
  Inferred information & Race and ethnicity & Race and ethnicity \\
  Linking attributes & Address/block, age, sex & Address/block \\
  Mechanism & Constraint-solver reconstruction & Simple table lookup \\
  Ground truth & Census' internal raw data & 2010 swap-protected release \\
  \midrule
  Precision/Recall & P=75\%, R=85\% (all) & P=75\%, R=98\% \\
     (block size) & P=92\%, R=17\%  (10-49) & P$>$=95\%, R=23\% \\
     & P=97\%, R=1.5\% (1-9) & P=100\%, R=11\% \\
  \bottomrule
\end{tabular}
  \caption{Summary of the Bureau's attack and our simple inference non-attack. The reported precision and recall for the re-identification attack are for all blocks, blocks with between 10 and 49 respondents, and blocks with between 1 and 9 respondents respectively. The reported precision and recall for the non-attack are limited to blocks where the majority race has at least 5 persons.
  }
  \label{tab:overview}
\end{table*}
% THE FOLLOWING IS HOW ABOWD DEFINES RECALL
%1-9: 4,093,151 confirmed of 308,745,538 total = 1.5\% recall, with 96.98\% precision
%10-49: 43,415,168 confirmed = 16\% recall, with 91.68\% precision
%50-99: 42,515,756 confirmed = 16.4\% recall, with 82\% precision
%100-249: 45,807,270 confirmed = 16.4\% recall, with 72.41\% precision
%250-499: 22,902,054 confirmed = 8.2\% recall, with 64.6\% precision
%500-999: 13,514,134 confirmed = 4.8\% recall, with 56.05\% precision
%1000+: 6,711,193 confirmed = 2.4\% recall, with 52.59\% precision
%total: 178,958,726 putative = 64\% recall, with 75.14\% precision
%CEF has 279,179,329 and this is in fact the number we use in table 6

% THE FOLLOWING IS HOW I DEFINE IT
%1-9: 4,220,571 putative of 279,179,329 total = 1.5\% recall, with 96.98\% precision
%10-49: 47,352,910 putative = 17\% recall, with 91.68\% precision
%50-99: 51,846,547 putative = 18.6\% recall, with 82\% precision
%100-249: 63,258,561 putative = 22.6\% recall, with 72.41\% precision
%250-499: 35,454,412 putative = 12.7\% recall, with 64.6\% precision
%500-999: 23,280,718 putative = 8.3\% recall, with 56.05\% precision
%1000+: 12,761,586 putative = 4.6\% recall, with 52.59\% precision
%total: 238,175,305 putative = 85\% recall, with 75.14\% precision
%CEF has 279,179,329 (prior knowledge records) and this is in fact the number we use in table 6

Our demonstration operates by merely predicting that the race/ethnicity of any given respondent is that of the majority race for the corresponding block\footnote{An idea borrowed from Ruggles et al.~\cite{ruggles2021role}.}. In Section~\ref{sec:violated}, we argue that it is in fact the intention of the Bureau that the majority race/ethnicity of any block can be accurately inferred. If this is so, then our demonstration is not an attack at all. Rather, it simply utilizes the statistical inferences that census data is supposed to enable. As such, we refer to our demonstration as a simple inference \emph{non-attack}.

In 2018, Ruggles et al.~\cite{ruggles2018implications} argued that the Bureau's reconstruction attack is not particularly effective, and that TDA is not necessary. In 2021, Ruggles and Van Riper~\cite{ruggles2021role} simulated a simple statistical random reconstruction from national-level statistics, and showed that it can be roughly as effective as the Bureau's reconstruction attack (see Section~\ref{sec:ruggles}).

The contribution of this paper is that it is a much more concrete demonstration of the mismeasure of the Bureau's attack, for two reasons. First, our non-attack uses the 2010 census data as high-quality ground truth. By contrast, Ruggles and Van Riper use simulated data. Second, our non-attack runs on the TDA-protected data itself. This demonstrates directly that either the Bureau does not intend to protect against this inference, or that TDA fails to provide the intended protections.

We argue that it is the former. Indeed it is important to note that the Bureau as far as we know has not run its own reconstruction attack against the TDA-protected release. We suspect that doing so would yield results similar to the same attack on the 2010 swap-protected release.

Finally, note that this paper is intentionally narrow in scope. It pertains only to inferring race/ethnicity. Other types of inferences (i.e. age) would not have similarly high precision. Likewise we say nothing about other reconstruction attacks that may exist and may be more effective than that demonstrated by the Bureau. Finally, we make no recommendations as to how the Bureau may better define its privacy measures and criteria.

%It is important to note that this does \textbf{NOT} mean that TDA is not private. Rather, it means that the Bureau's criteria for measuring privacy leads to an \emph{incorrect interpretation} of the privacy implications of their re-identification attack. The Bureau's interpretation is overly pessimistic, and should not be used to as evidence that the 2010 census needs stronger protections.

%One way to understand the Bureau's mistake is with the following analogy. Suppose we wanted to measure deaths due to COVID in 2021. One approach might be to simply count all deaths in 2021 and attribute them to COVID. This is obviously nonsense, since most people will have died for other reasons. A better approach would be to compute the expected number of deaths in 2021 by looking at prior years, and assume that the deaths \emph{in excess of this baseline} in 2021 are due to COVID.

%The Bureau failed to take into account any kind of \emph{privacy baseline}, specifically the intentional and legitimate release of statistical information about groups of individuals. Our non-attack demonstrates this by showing that the inference of race and ethnicity on the TDA-protected release exceeds that of their own attack.

Section~\ref{sec:bureau} describes the Bureau's re-identification attack in more detail. Section~\ref{sec:inference} describes our non-attack. Section~\ref{sec:conclude} explores the question of whether our non-attack represents a meaningful privacy loss or (more likely) not.

The code and data for our non-attack may be found at
\ifblind
(URL blinded. Files available upon request.)
\else
\href{https://gitlab.mpi-sws.org/francis/census-misinterpretation}{https://gitlab.mpi-sws.org/francis/census-misinterpretation}.
\fi

\section{The Bureau's Re-identification Attack}
\label{sec:bureau}

The Bureau's re-identification attack, as well as our non-attack, combine \emph{prior knowledge} with \emph{released data} to infer the race and ethnicity of a \emph{target} individual (the census respondent). Table~\ref{tab:overview} summarizes both attacks.

\begin{figure}
\begin{center}
\includegraphics[width=0.5\linewidth]{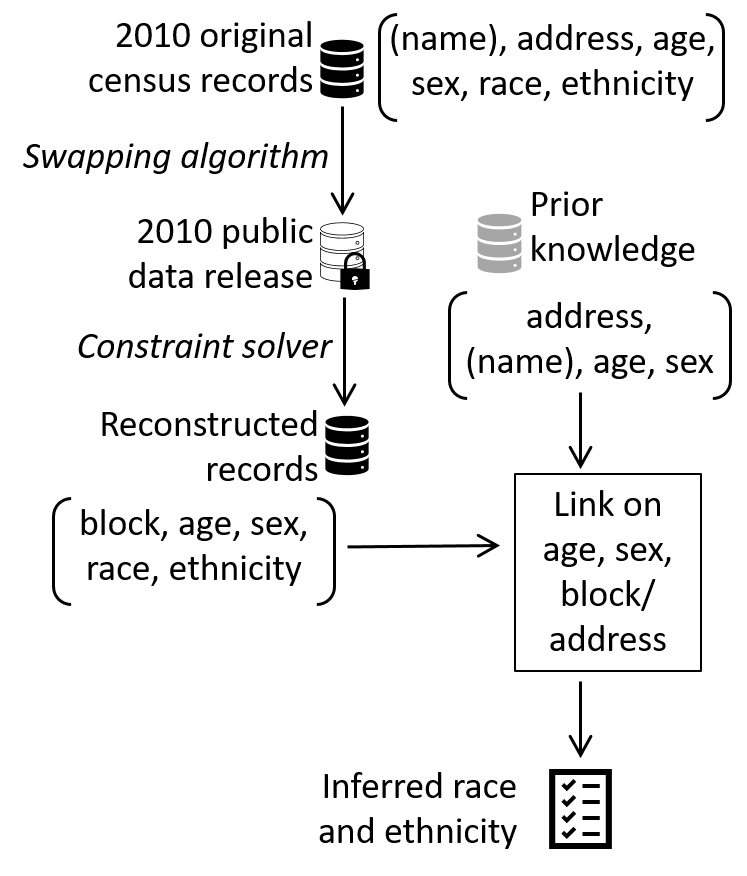}
\caption{Bureau's re-identification attack.}
\label{fig:census}
\end{center}
\end{figure}

The re-identification attack works as follows (Figure~\ref{fig:census}):
\begin{enumerate}
    \item Derive a set of constraints from the swap-protected release (per block).
    \item Use a constraint solver to reconstruct the original data per block. The resulting reconstructed records are pseudonymous (not linked to identified persons).
    \item Link the reconstructed records to externally-derived prior knowledge data on attributes shared by the reconstructed data and the prior knowledge data. This serves to identify the persons in the reconstructed data.
    \item Infer the unknown attributes from the so-identified reconstructed data.
\end{enumerate}

A good overview of the re-identification attack and its results can be found in Abowd~\cite{fair-lines-lawsuit}. A general description of the constraint solver approach can be found in Garfinkel et al.~\cite{garfinkel2018understanding}.

In the attack demonstrated by the Bureau, the reconstructed data consisted of attributes \emph{block, age, sex, race, and ethnicity}. Blocks are geographical areas ranging from zero persons to several thousand persons. There are 6M blocks in the USA. There are two ethnicity values, \emph{Hispanic} and \emph{Not Hispanic}. There are 63 race values. The values are built from six basic categories (\emph{white, black, asian, native, island, and other}), either individually or in combinations (mixed race). The majority of race values, however, are white, black, or asain. In total there are 126 race/ethnicity combinations.

The Bureau ran the attack twice using two different sources of prior knowledge. One source consists of commercially-available data, and can therefore be run by anybody. The accuracy of the commercial data is questionable, leading to some uncertainty as to whether any lack of attack effectiveness is due to errors in reconstruction or errors in the prior knowledge. The second source consists of the Bureau's own internal data, and is therefore a perfect match. This represents the worst-case scenario. For the purpose of comparing our non-attack with the Bureau's re-identification attack, we focus only on these worst-case results. This second internal source is referred to as CEF (Census Edited File) in Abowd\cite{fair-lines-lawsuit}.

A reconstructed record is correct when its name, address, age (within one year), sex, race, and ethnicity match with a record in the Bureau's internal data.

Abowd\cite{fair-lines-lawsuit} provides measures of \emph{precision} and \emph{recall}.  Precision is defined as the percentage of correct records to linked records, and we use this definition as well (i.e. in Table~\ref{tab:overview} and Figure~\ref{fig:precision_per_bin}). Abowd's recall is defined as the percentage of correct reconstructed records to all prior knowledge records. This definition doesn't make sense to us because the attacker does not know whether a reconstructed record is correct or not, and therefore has no basis on whether to make a prediction or not. We therefore use a different measure of recall: the fraction of linked records to total prior knowledge records. This definition is used in Table~\ref{tab:overview}.

% we have a set of prior knowledge records
% some fraction don't match reconstructed records: these should define recall
% some fraction of matched records are incorrect: these should define precision

There are 279,179,329 prior knowledge records (Abowd's Table 2, column \emph{``Records with PIK, Block, Sex, and Age''}, row \emph{``CEF''}). Abowd's Table 6 gives the number of linked records (column \emph{``Putative Re-identifications (Source: CEF)''}) and the number of correct records (column \emph{``Confirmed Re-identifications (Source CEF)''}) for each binned block size as well as for all blocks taken together. Abowd's Table 6 provides the precision measures. The recall measures can be computed by dividing the putative re-identifications by the total prior knowledge records.

Three of the precision and recall values are given in (this paper's) Table~\ref{tab:overview}. One is for all blocks, and the other two are for the block sizes with the highest precision.

\section{Our Simple Inference Non-attack}
\label{sec:inference}

Our inference mechanism (non-attack) requires only the street address of the target individual as prior knowledge. The information needed to link street address to census block is public information\footnote{For instance \href{https://geocoding.geo.census.gov}{https://geocoding.geo.census.gov}}.

To infer a person's race and ethnicity, we simply look up the \emph{majority} race and ethnicity for the person's census block, and infer that this is the race/ethnicity of the person. The majority race/ethnicity for a given block is that with the highest count of all 126 race/ethnicity values.

We define \emph{precision} as the majority race/ethnicity count divided by the block count. This is simply the statistical probability that any given person in a block indeed has the majority race/ethnicity.

To measure precision, we use the 2010 swap-protected release as the ground truth. This is not a perfect ground truth, but we believe that it is close enough for our purpose, which is merely to show that the Bureau's re-identification attack and our simple inference non-attack are in the same ballpark.

In the 2010 swap-protected release, the block count is an exact count (not noisy, see paragraph 37 of Abowd\cite{fair-lines-lawsuit}). Because of swapping, race/ethnicity counts may not be exact, but we assume that the majority race/ethnicity count is almost always exact. This is because swapping generally occurs with households that are unique, i.e. those with rarer race/ethnicities. Our non-attack depends only on the count of the majority race/ethnicity, which is less prone to distortion from swapping. Note that the exact parameters for swapping are not published~\cite{census2019legacy} so we are not completely certain of this assumption.

The block-level 2010 swap-protected and TDA-protected releases were compiled into a single table\footnote{\href{https://www.nhgis.org/privacy-protected-2010-census-demonstration-data}{https://www.nhgis.org/privacy-protected-2010-census-demonstration-data}, Vintage 2021-06-08.} by IPUM NHGIS (National Historical Geographic Information System). The block-level data is available as per-state tables, so we merged the tables for all states plus DC and Puerto Rico for our measure. We measured only voting-age (over 18 years) counts to better compare with the Bureau's attack, which also uses voting age data (paragraph 5 from Abowd\cite{fair-lines-lawsuit}).

Our procedure for measuring per-block precision goes as follows:

\begin{enumerate}
\item Find the block's majority race/ethnicity $MR_{TDA}$ from the TDA-protected data (the race/ethnicity with the highest count).
\item Set the majority count $MC_{GT}$ as the count for race/ethnicity $MR_{TDA}$ from the swap-protected data (ground truth).
\item If $MC_{GT} < 5$, set precision as zero (see Section~\ref{sec:threshold}).
\item Otherwise, set the block count $BC_{GT}$ as that of the swap-protected data (ground truth).
\item Set precision as $MC_{GT}/BC_{GT}$.
\end{enumerate}

In most cases, the TDA-protected data and the swap-protected data have the same majority race/ethnicity ($MR_{TDA} = MR_{SWAP}$). In this case, there is no precision penalty incurred by the noise from TDA. The majority race/ethnicity is different in blocks where the majority race/ethnicity and the second race/ethnicity have similar counts. In these cases, the noise from TDA can be enough to promote the second race/ethnicity to the majority. The precision penalty in these cases, however, is relatively small since the counts are not very different.

For example, suppose that the ground-truth count for the majority race/ethnicity for a block with 100 persons is 51, and for the second race/ethnicity is 49. However, the noise from TDA causes the second race/ethnicity to be the majority. In this case, the measured precision for TDA will be 49\% instead of the correct 51\% --- the precision penalty paid for the noise is relatively small.

The \emph{recall} for a given precision is measured as the fraction of records with the given precision or better for blocks where there are at least 5 persons with the majority race/ethnicity. Table~\ref{tab:overview} gives several illustrative precision/recall values.

Figure~\ref{fig:precision_cdf} is a CDF showing the precision of all records for our non-attack executed on both the TDA-protected and swap-protected releases. A precision of zero is conservatively assigned when the associated block has fewer than 5 persons with the majority race. The precision of the non-attack on the swap-protected data is virtually identical to that of the TDA-protected data. In other words, the TDA protection does not affect the ability to (statistically) infer peoples' race and ethnicity.

Also shown in Figure~\ref{fig:precision_cdf} are the precision measures taken from Abowd\cite{fair-lines-lawsuit}. Each point represents a different range of block sizes. Prior-knowledge records without a match among the reconstructed records are assigned a precision of zero. From Figure~\ref{fig:precision_cdf} we see that simple inference is substantially more effective than the Bureau's re-identification.

Note that in our non-attack, the attacker knows roughly what precision any given block has. Because of noise added to TDA-protected counts, an attacker does not know the exact precision for a given block. The amount of noise, however, is relatively small, so the attacker has a good estimate of precision.

The race/ethnicity of a substantial fraction of the population (11\%) can be inferred with virtually 100\% precision. This is for the simple reason that many blocks have only a single race/ethnicity.

Note that this result is supported by Kenny et al.~\cite{kenny2021use}, which also predicts individual race and ethnicity, but additionally uses analysis of names to help predict race and ethnicity. Kenny et al. found that TDA (using a version of TDA with more noise than the final version we tested) did not degrade the quality of these predictions.

\begin{figure}
\begin{center}
\includegraphics[width=0.7\linewidth]{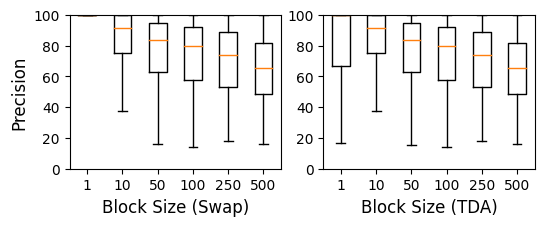}
\caption{Non-attack precision by block size for TDA-protected and swap-protected releases.}
\label{fig:precision_per_bin}
\end{center}
\end{figure}

Figure~\ref{fig:precision_per_bin} shows the non-attack precision as a whisker plot per block size group (1-9 persons, 10-49 persons etc.). This figure shows that the noise of TDA substantially distorts the data for the smallest blocks (9 or fewer persons), but not for blocks larger than that. We assume that the near-perfect precision measure for the swap-protected data for the smallest blocks is because small blocks tend to be very homogeneous, and because swapping removes most of what little non-homogeneity remains. Note in particular that the relative distortion due to swapping is greater for small blocks. As such, the near-perfect precision measure for small blocks for swap-protected data does not accurately reflect reality. In other words, it would not necessarily be correct to conclude from this figure that swapping fails to protect privacy for small blocks.

\begin{figure}
\begin{center}
\includegraphics[width=0.6\linewidth]{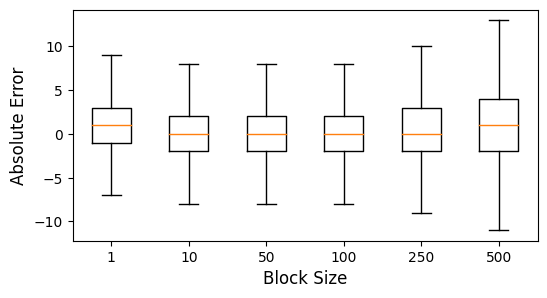}
\caption{Absolute error in count introduced by TDA noise between swap-protected and TDA-protected data for the same race/ethnicity in each block. Measured as TDA-count minus swap-count.}
\label{fig:error_per_bin}
\end{center}
\end{figure}

Figure~\ref{fig:error_per_bin} shows the absolute error between swap-protected and TDA-protected counts. Most of the comparisons are for the majority race/ethnicity. When the two releases have different majority race/ethnicity, the comparison uses the majority race of the TDA-protected data. The error is relatively small, less than plus or minus 10 in most cases. The standard deviation in error for the smallest blocks is 2.8, and 4.0 for the largest blocks. This supports the observation that most of the loss of precision occurs for the smallest blocks. We don't know enough about TDA or swapping to understand why the median error is greater than zero, or why the range of error increases with larger blocks.

\subsection{Effect of majority race/ethnicity threshold}
\label{sec:threshold}

In our non-attack, we ignore blocks where the majority race/ethnicity has fewer than 5 persons. The idea here is that a group of 5 persons is a reasonable privacy threshold. In other words, revealing that there are at least 5 persons with a given race and ethnicity does not unduly compromise the privacy of those 5 persons since they are not race/ethnicity uniques. The measures given in Table~\ref{tab:overview} and Figure~\ref{fig:precision_cdf} use this threshold.

\begin{figure}
\begin{center}
\includegraphics[width=0.7\linewidth]{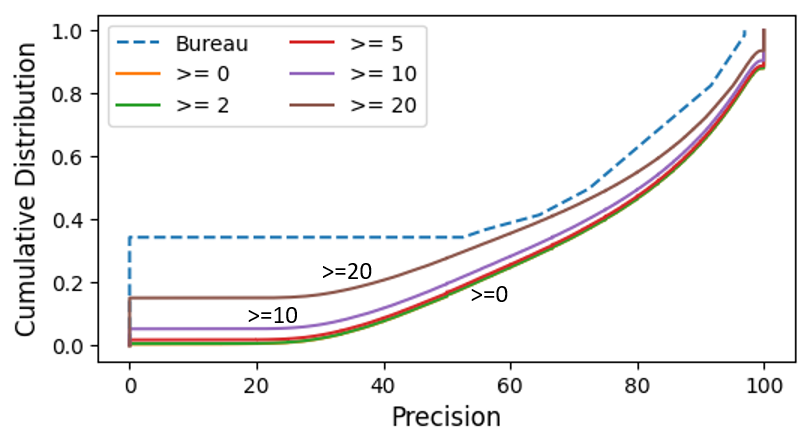}
\caption{CDF of precision for our non-attack for different thresholds for the number of persons with the majority race/ethnicity. Blocks with fewer than the threshold number of persons are conservatively assigned a precision of zero. For comparison, the precision for the Bureau's re-identification attack on the prior swap mechanism is also shown.}
\label{fig:thresholds}
\end{center}
\end{figure}

The choice of 5 is our own, and others may feel that a larger threshold is required, or that a smaller threshold is adequate. We therefore give the cumulative distribution of precision for several different thresholds in Figure~\ref{fig:thresholds}. Note that even with a threshold of 20, a substantial fraction of blocks allow correct inference with 100\% precision.

\section{Discussion and Conclusion}
\label{sec:conclude}

In this paper, we show that the TopDown Algorithm (TDA) used by the US Census Bureau for the 2020 census in no way prevents attackers from inferring race and ethnicity with high accuracy for a substantial portion of census respondents.

Here we discuss two questions. First, does the ability to make this inference constitute a privacy violation of some sort? Note that it not up to us to answer this question. Rather, this is a matter for US Census and US Government policy makers. Nevertheless, if the answer is no, then the US Census has incorrectly measured the effectiveness of their attack. If the answer is yes, then TDA is an inadequate defense.

Second, assuming the answer is no (there is no privacy violation), then what did the Bureau do wrong in its privacy measure?

\subsection{Has privacy been violated?}
\label{sec:violated}

The US Census Bureau has quite clearly stated that it is a goal to allow for accurate statistical inferences. The following is from~\cite{census-disclosure-avoidance}:

\begin{quote}
\emph{Some inferences about confidential information can be achieved with purely statistical information (especially for blocks with many identical records). These inferences rely on aggregate statistical information about groups and do not rely on any individuals' confidential census responses. For example, suppose Alice is trying to learn how Bob responded to the race question, and she already knows Bob lived in Montana at the time of the 2010 Census enumeration.  Alice could then review the 2010 Census tables, and because she can find that 89.4 percent of respondents reported ``White Alone'' in Montana, Alice can guess with high confidence that Bob's census response was ``White Alone.'' This is an example of an inference based on aggregate statistical information about groups, rather than knowledge of Bob's confidential census response. The Disclosure Avoidance System (DAS) permits accurate inferences based on aggregate statistical information about groups.}
\end{quote}

It seems to us very clear from this that the US Census Bureau intends to allow accurate statistical inferences. Indeed the Census here has for all practical purposes described our non-attack. As with our non-attack, Alice's best strategy is to simply guess the majority attribute (here race). The only differences are that we use address mapped to block rather than state, and infer both race and ethnicity.

\subsection{What is wrong with the US Census privacy measure?}

It is important to point out that the Bureau doesn't use inference as its measure of privacy. In Abowd\cite{fair-lines-lawsuit}, two measures of privacy are given, neither of which is inference.

The first measure is simply the fraction of correctly reconstructed records. Referring to Figure~\ref{fig:census}, this would be the fraction of reconstructed records that match the original census records on block, age, sex, race, and ethnicity.

The first seven pages of Abowd\cite{fair-lines-lawsuit} pertain to the reconstruction measure. These seven pages conclude with:

\begin{quote}
\emph{Consequently, the new technology-enabled possibility of accurately reconstructing HDF microdata from the published tabular summaries and the fact that those reconstructed data do not meet the disclosure avoidance standards established at the time for microdata products derived from the HDF demonstrate that the swapping methodology as implemented for the 2010 Census no longer meets the acceptable disclosure risk standards established when that swapping mechanism was selected for the 2010 Census.}
\end{quote}

In other words, the mere ability to reconstruct with some level of success (see Abowd's Table 1), \emph{whether or not the reconstructed records can be matched to named persons}, is the \emph{only} criteria required by the Bureau.

A simple thought experiment, however, shows that reconstruction alone is not a valid measure of privacy. Imagine, for instance, a table with one column, sex, with values `M' and `F' over thousands of rows. Given noisy counts of the two values, it would be easy to reconstruct the table with high accuracy. Clearly this in and of itself is not a violation of privacy. At a minimum, the number of unique values in the table is also important. The census data used for the attack has only 57\% uniques on the five attributes.

In any event, the Bureau recognizes the importance of being able to link reconstructed records with named persons, and so the subsequent 12 pages of Abowd\cite{fair-lines-lawsuit} focus on the second measure, that of re-identification. This measures the fraction of reconstructed records that can be successfully linked to prior knowledge of named persons. Referring again to Figure~\ref{fig:census}, this is the box labeled ``Link on age, sex, block/address''.

Strictly speaking, this re-identification measures the ability to identify name, address, age, sex, race, and ethnicity with some level of success. Abowd\cite{fair-lines-lawsuit} reports that the Bureau finds that this also fails its criteria for privacy:

\begin{quote}
\emph{The Data Stewardship Executive Policy Committee (DSEP) determined that the simulated attack success rates in Table 6 were unacceptable for the 2020 Census. Decennial census data protected by the 2010 disclosure avoidance software is no longer safe to release.}
\end{quote}

Although Abowd\cite{fair-lines-lawsuit} never uses the word `infer', given that re-identification requires prior knowledge of name, age, sex, and address, it seems perfectly reasonable to describe the re-identification as inferring race and ethnicity from name, age, sex, and address. The Bureau is implicitly saying that this is not acceptable. On the other hand, they are saying that it is ok to infer race and ethnicity from address, because they very intentionally release data designed to do this.

There are perhaps three possible responses that the Bureau could make to this apparent contradiction:

\begin{enumerate}
    \item Re-identification using prior knowledge and inference using the same or less prior knowledge are different and can't be compared.
    \item It is not this specific re-identification per se that is a problem, but the fact that re-identification can happen in general.
    \item The re-identification doesn't really matter, since in any event the reconstruction alone failed the privacy criteria.
\end{enumerate}

The first seems non-sensical.

Regarding the second, it would be helpful if the Bureau identified cases of re-identification that revealed substantially more information than what the data release is supposed to reveal statistically.

\subsection{The Ruggles and Van Riper reconstruction}
\label{sec:ruggles}

Regarding the third, Ruggles and Van Riper~\cite{ruggles2021role} provide evidence that even the Bureau's reconstruction reveals nothing more than what is meant to be revealed statistically. In their demonstration, Ruggles et al. used the following statistical information as the basis for reconstruction (taken from the 2010 census):

\begin{enumerate}
\item The national distributions of ages, sexes, and block sizes.
\item The fact that 78\% of individuals on average have the majority race/ethnicity of their block.
\end{enumerate}

Armed with only this knowledge, Ruggles et al. built 10000 synthetic blocks by randomly assigning block size according to the national distribution, and then randomly assigning individuals to the blocks with age and sex following the national distributions. They then mimicked reconstruction of the synthetic blocks by fresh random assignments age and sex following the national distributions. Finally, they assumed that on average the race and ethnicity would be correct 78\% of the time.

The result is that on average 41\% of records matched on all five attributes (block, age, sex, race, and ethnicity), compared to 45\% for the Bureau's attack (using commercially obtained name, age, sex, and address). These two reconstruction measures are certainly in the same ballpark.

The idea that the Bureau's reconstruction is little better than random is also supported by Muralidhar~\cite{muralidhar2022reexamination}. He shows that the Bureau's reconstruction can produce a large number of different solutions. Any given solution chosen by the Bureau is effectively a random choice among many.

\subsection{Conclusion}

In conclusion, we believe that the Bureau has not adequately demonstrated a meaningful privacy threat against the 2010 swapping method. The threat may exist, but has not been demonstrated. We also believe that the criteria used by the Bureau to measure privacy is flawed in that it does not take the intended released statistical knowledge into account. This is demonstrated partially by the Ruggles and Van Riper reconstruction, and more definitively by the inference non-attack of this paper.

Note that implementing TDA has been costly both in terms of data quality and timely data release. Numerous studies point to problems for a variety of research tasks and government functions, including redistricting~\cite{kenny2021use}, health~\cite{santos2020differential}\cite{hauer2021differential}\cite{santos2021changes}, and demographics~\cite{santos2020differential}\cite{mueller20212020}\cite{winkler2021differential}. (Note that some of these studies may be based on earlier proposed versions of anonymization with more noise than the final version.)

The state of Alabama filed a (failed) lawsuit in part to force the Census Bureau return to the former low-distortion method of anonymization~\cite{lawsuit-alabama}, and a second lawsuit to force the Bureau to release delayed housing data is ongoing as of this writing (Spring 2022)~\cite{lawsuit-fairlines}. The Bureau has yet to release all of the tables that it normally releases.

Although this paper does not make any concrete proposals on how better to measure privacy, it seems clear to us that more research is needed, especially regarding the role that expected statistical inference plays in measuring privacy. We hope that this paper serves to motivate that research, and that it leads to a more circumspect approach to measuring privacy loss in statistics organizations.

\bibliographystyle{abbrv}
\bibliography{../../masterBib/master}

\end{document}